\documentclass{aastex}
\usepackage{emulateapj5}
\usepackage{apjfonts}

\usepackage{epsf}

\slugcomment{Accepted for publication in the Astrophysical Journal}

\newcommand{\pfrac}[2]{{\left(\frac{#1}{#2}\right)}}

\newcommand{\gdot}{{\dot{\gamma}}}
\newcommand{\e}{{\epsilon}}
\newcommand{\et}{{\tilde{\epsilon}}}
\newcommand{\emax}{{\epsilon_{\rm max}}}
\newcommand{\jsyn}{{j_{\rm syn}}}
\newcommand{\jssc}{{j_{\rm ssc}}}
\newcommand{\usyn}{{u_{\rm syn}}}
\newcommand{\rblr}{{r_{\rm BLR}}}

\begin{document}

\title{Synchrotron and Synchrotron Self-Compton Spectral Signatures
and Blazar Emission Models}

\author{James Chiang\altaffilmark{1,2} and Markus B\"ottcher\altaffilmark{3,4}}
\altaffiltext{1}{NASA/GSFC, Code 661, Greenbelt MD 20771}
\altaffiltext{2}{Joint Center for Astrophysics/Physics Department, 
                 University of Maryland, Baltimore County, Baltimore
                 MD 21250}
\altaffiltext{3}{Department of Physics and Astronomy, 
                 Rice University, Houston TX 77005-1892}
\altaffiltext{4}{Chandra Fellow}

\begin{abstract}
We find that energy losses due to synchrotron self-Compton (SSC)
emission in blazar jets can produce distinctive signatures in the
time-averaged synchrotron and SSC spectra of these objects.  For a
fairly broad range of particle injection distributions, SSC-loss
dominated synchrotron emission exhibits a spectral dependence $F_\nu
\sim \nu^{-3/2}$.  The presence or absence of this dependence in the
optical and ultraviolet spectra of flat spectrum radio quasars such as
3C~279 and in the soft X-ray spectra of high frequency BL Lac objects
such as Mrk 501 gives a robust measure of the importance of SSC
losses.  Furthermore, for partially cooled particle distributions,
spectral breaks of varying sizes can appear in the synchrotron and SSC
spectra and will be related to the spectral indices of the emission
below the break.  These spectral signatures place constraints on the
size scale and the non-thermal particle content of the emitting plasma
as well as the observer orientation relative to the jet axis.
\end{abstract}

\keywords{galaxies: active --- gamma rays: theory --- radiation
          mechanisms: non-thermal}

\section{Introduction}

A central uncertainty in the modeling of high energy radiation from
blazar jets is the identity of the primary energy loss mechanism which
controls the energy balance of the non-thermal plasma believed to
produce the observed radiation (e.g., Sikora et al.\ 1997).  Because
of ambiguities in observer orientation relative to the jet axis and
differing Doppler beaming patterns for the various radiation
processes, direct measurement and simple modeling of the high energy
spectral components cannot provide conclusive results.  In this
paper, we describe spectral signatures which may appear in the
time-averaged synchrotron and synchrotron self-Compton (SSC) spectra
of these objects. If they are present, then they provide strong
evidence for the dominance of energy losses due to synchrotron
self-Compton emission rather than other processes such as external
Compton scattering or synchrotron emission.

\section{Synchrotron Self-Compton Cooling and Time-Averaged Spectra}

We consider the cooling of non-thermal electrons in a blob of plasma
which has bulk Lorentz factor $\Gamma$.  Following the blazar model of
B\"ottcher, Mause, \& Schlickeiser (1997), we assume that this blob
starts off at a height $z_i$ above the accretion disk, and travels
outward along the jet axis.  The non-thermal electrons and/or
positrons which produce the observed synchrotron and inverse Compton
emission are assumed to be injected at $z_i$ with an energy
distribution
\begin{equation}
\frac{dn}{d\gamma_i} = n_0 \gamma_i^{-p},
                       \qquad \gamma_{1,i} \le \gamma_i \le \gamma_{2,i},
\end{equation}
where $\gamma_i$ is the Lorentz factor of an individual particle,
$\gamma_{2,i}$ and $\gamma_{1,i}$ are the maximum and minimum values
at the time of injection, and $n_0$ is related to the total particle
density $n_{\rm tot}$ by $n_0 = n_{\rm tot} (1-p)/(\gamma_{2,i}^{1-p}
- \gamma_{1,i}^{1-p})$.

For synchrotron cooling and inverse Compton cooling in the Thomson
limit, the single particle energy loss rate is given by $\gdot =
-\beta (\gamma^2-1)$ where $\beta = (4\sigma_T/3 m_e c)(u_B + u_{\rm
rad})$, $m_e$ is the electron mass, $\sigma_T$ is the Thomson
cross-section, $u_B = B^2/8\pi$ is the magnetic field energy density,
and $u_{\rm rad}$ is the ambient photon energy density in the blob
frame.  If $\beta$ is constant, then for $\gamma \gg 1$, the energy of
an individual particle in the co-moving frame of the blob evolves as
\begin{equation}
\gamma = \frac{\gamma_i}{1 + \gamma_i\beta t},
\label{gamma(t)}
\end{equation}
and the particle distribution as a function of time is
\begin{equation}
\frac{dn}{d\gamma} = \frac{n_0 \gamma^{-p}}{(1 - \gamma\beta t)^{2-p}}.
\end{equation}
For injection indices $p \simeq 2$, the {\em shape} of the particle
spectrum remains roughly constant, becoming somewhat softer with time
at the highest particle energies for $p > 2$ and somewhat harder for
$p < 2$.  The maximum and minimum Lorentz factors, $\gamma_2$ and
$\gamma_1$, evolve according to Eq.~\ref{gamma(t)}, so that the entire
particle spectrum marches downward in energy as time passes.

To estimate the time-averaged synchrotron spectrum, we integrate the
$\delta$-function approximation of Dermer \& Schlickeiser (1993;
hereafter DS93) for the instantaneous synchrotron spectral emissivity:
\begin{equation}
\jsyn(\e, t) = \frac{c\sigma_T}{6\pi\e_B} u_B \pfrac{\e}{\e_B}^{1/2} 
              \frac{dn}{d\gamma}\left[\pfrac{\e}{\e_B}^{1/2}\right],
\label{jsyn(t)}
\end{equation}
where $\e = h\nu/m_ec^2$, $\e_B = B/B_{\rm crit}$, and $B_{\rm crit}
\equiv m_e^2c^3/e\hbar = 4.4\times 10^{13}$\,G.  The time-averaged
spectrum is then
\begin{eqnarray}
\jsyn(\e) 
      & =      & \frac{1}{t} \int_0^t \jsyn(\e, \tilde{t})d\tilde{t} \\
      & \simeq & \jsyn(\e, 0)\, \min(1, t_{\rm max}/t),
        \label{jsyn_approx}
\end{eqnarray}
where
\begin{equation}
t_{\rm max}(\e) = \frac{1}{\beta}\left[\pfrac{\e_B}{\e}^{1/2} 
                                       - \frac{1}{\gamma_{2,i}}\right]
\label{tmax}
\end{equation}
is the time during which the highest energy particles emit synchrotron
radiation with characteristic energy $\emax = \e_B\gamma_2^2(t) > \e$.
The approximate relation given by Eq.~\ref{jsyn_approx} arises because
of the near constant shape of the particle distribution for $p \sim
2$.  Combining Eqs.~\ref{jsyn_approx} and \ref{tmax}, we recover the
standard result for synchrotron emission from a partially cooled
particle distribution:
\begin{eqnarray}
\jsyn(\e) &\sim &\e^{(1-p)/2}, \qquad\qquad\qquad \e < \emax \\
          &\sim &\e^{(1-p)/2 - 1/2} \sim \e^{-p/2}, \qquad \emax < \e.
           \label{jsyn_constant_beta}
\end{eqnarray}
This result agrees with the treatment of Dermer, Sturner, \&
Schlickeiser (1997; their Eq.~55) who considered the case of
synchrotron losses and losses due to Thomson scattering of external
photons.  Both the present derivation of Eq.~\ref{jsyn_constant_beta}
and that of Dermer et al.\ (1997) formally rely on $\beta$ being
constant.  It will still apply if $\beta$ changes on time scales much
longer than the typical energy loss time scale.  For conical expansion
of the jet outflow, magnetic flux freezing implies $B(z) = B_0
(z/z_i)^{-\delta}$, where the index $\delta$ is of order unity.  If
synchrotron emission is the primary loss mechanism, then
Eq.~\ref{jsyn_constant_beta} will hold if
\begin{equation}
z_i \gg \frac{6\pi m_e c^2\delta\Gamma}{\sigma_T \gamma B_0^2} 
    \sim 10^{16}\,{\rm cm} \pfrac{10^5}{\gamma} \pfrac{\Gamma}{10}
         \pfrac{1\,{\rm G}}{B_0}^2
\end{equation}
(cf.\ DS93).  Since limits on X-ray emission due to bulk
Comptonization imply injection heights of $z_i \gg 200\,r_g$ where
$r_g \equiv GM/c^2 \sim 10^{13}$\,cm for a $10^8$\,M$_\odot$ black
hole (Sikora et al.\ 1997), we can assume the magnetic field to be
effectively constant.

Energy losses due to first order synchrotron self-Compton (SSC)
emission can be estimated by using the synchrotron photon energy
density:
\begin{eqnarray}
\usyn &\simeq &\frac{4}{3} c\sigma_T u_B \pfrac{3R_b}{4c}
            \int_{\gamma_1}^{\gamma_2} \frac{dn}{d\gamma}\gamma^2 d\gamma \\
      &\simeq & \sigma_T u_B R_b \frac{n_0}{3-p} 
                (\gamma_2^{3-p} - \gamma_1^{3-p}), \qquad p \ne 3.
\label{usyn}
\end{eqnarray}
Here $3R_b/4c$ is the mean residence time for photons produced in a
spherical blob of radius $R_b$.  Assuming $\usyn > u_B$, $\gamma_2 \gg
\gamma_1$, and $p \la 3$, the single particle energy loss rate is
\begin{equation}
\gdot \simeq -\frac{4}{3}\frac{\sigma_T^2}{m_e c} R_b u_B
             \frac{n_0}{3-p} \gamma_2^{3-p}\gamma^2,
\label{gdot_ssc}
\end{equation}
and the maximum particle energy evolves according to
\begin{equation}
\gdot_2 \simeq -\tilde\beta\gamma_2^{5-p},
\label{gdot2_ssc}
\end{equation}
where $\tilde\beta \equiv (4\sigma_T^2/3 m_e c) R_b u_B n_0/(3-p)$.  In
analogy to Eqs.~\ref{jsyn_approx} and \ref{tmax}, this implies
\begin{equation}
t_{\rm max}(\e) = \frac{1}{(4-p)\tilde\beta}\left[\pfrac{\e_B}{\e}^{(4-p)/2}
                                  - \frac{1}{\gamma_{2,i}^{4-p}}\right],
\label{tmax_ssc}
\end{equation}
and for the time-averaged synchrotron spectrum, we have
\begin{equation}
\jsyn(\e) \sim \e^{(1-p)/2 - (4-p)/2} \sim \e^{-3/2}, \qquad \emax < \e.
           \label{jsyn_final}
\end{equation}
Here again $\emax = \e_B \gamma_2^2$, but in this case, $\gamma_2$
evolves according to Eq.~\ref{gdot2_ssc}.  Notably, the $p$-dependence
drops out of the time-averaged spectrum for the highest energy
synchrotron photons.

The time-averaged SSC spectrum can be estimated in a similar fashion.
For this purpose, we use a similar $\delta$-function approximation for
the SSC emissivity (DS93):
\begin{eqnarray}
\jssc(\e, t) &\simeq& \frac{c\sigma_t^2 u_B R_b}{9\pi\e_B^{1/2}}\pfrac{\e}{\e_B}
              \int_{\e_B\gamma_1^2}^{\min(\e_B\gamma_2^2, \e, 1/\e)}
               d\et \et^{-1} \nonumber\\
            &\times& \frac{dn}{d\gamma}\left[\pfrac{\et}{\e_B}^{1/2}
               \right] \frac{dn}{d\gamma}\left[\pfrac{\e}{\et}^{1/2}\right].
\label{jssc(e,t)}
\end{eqnarray}
This expression shows that the instantaneous SSC spectrum has the same
spectral index, $\alpha = (p-1)/2$ (where $F_\nu \sim \nu^{-\alpha}$),
as the synchrotron component (Eq.~\ref{jsyn(t)}).  However, the
maximum photon energy is instead related to the maximum electron
Lorentz factor by $\emax = \min(\gamma_2, \e_B\gamma_2^4)$ where the
former value holds for early times, but only very briefly for typical
blazar parameters (see Fig.~\ref{example_seds}).  Modifying
Eq.~\ref{tmax_ssc} accordingly, this implies for the high energy
portion of the time-averaged SSC spectrum
\begin{equation}
\jssc(\e) \sim \e^{(1-p)/2 - (4-p)/4} \sim \e^{-(2+p)/4}.
\end{equation}
Even though the instantaneous synchrotron and SSC spectra have similar
spectral indices, the time-averaged index of the SSC emission will be
harder than that of the synchrotron emission for $p \la 3$.  As a
corollary to this, if synchrotron emission is the most important
cooling process, the time-averaged SSC spectrum goes as $\jssc \sim
\e^{(1-p)/2 - 1/4} \sim \e^{(1-2p)/4}$
(cf. Eq.~\ref{jsyn_constant_beta}).

The preceding discussion assumes that the shape of the particle
distribution retains the same power-law index $p$ as it evolves.
Since the energy loss rates for all the electrons (except the highest
energy ones) still have a $\gamma^2$-dependence (Eq.~\ref{gdot_ssc}),
this will continue to be approximately true for injection indices $p
\sim 2$.  Therefore, if SSC losses dominate the particle cooling, the
synchrotron emission above the break energy $\emax$ will have a
time-averaged spectral shape of $F_\nu \sim \nu^{-3/2}$ and will be
relatively insensitive to the shape of the injected particle
distribution.  As we noted above, for $p > 2$ the particle
distribution becomes softer with time, while for $p < 2$ it becomes
harder.  Thus we would expect to see a distribution of spectral
indices in the optical/UV clustered around $\alpha = 3/2$ as the
injection index varies.  In the Thomson limit for SSC losses,
Eq.~\ref{jsyn_final} is a surprisingly good approximation over a broad
range of injection indices.  In Fig.~\ref{example_seds}, we plot the
time-averaged $\nu F_\nu$ synchrotron and SSC spectra for injection
indices $p = 0.5$--3.5.  These spectra have been computed by
numerically integrating the full expression for particle energy loss
rates including synchrotron and SSC processes.  Here we have used the
formulae of Crusius \& Schlickeiser (1986, 1988) to compute the
instantaneous synchrotron spectra since they give more accurate
results than Eq.~\ref{jsyn(t)}.  For $p = 3$, which is at the limit of
validity for Eq.~\ref{gdot_ssc}, the time-averaged synchrotron
spectral index is $\alpha \simeq 1.6$.  Even for $p = 3.5$, we find
$\alpha \simeq 1.8$.  If one takes the Klein-Nishina roll-over into
account, then the SSC energy loss rate will deviate from the
Thomson-limit expression for the highest energy electrons (B\"ottcher
et al.\ 1997).  This causes a flattening of the time-averaged
synchrotron spectrum at the highest photon energies.  The thick curve
in Fig.~\ref{example_seds} is the time-averaged synchrotron spectrum
for a $p = 2$ calculation which uses the full Klein-Nishina
cross-section in determining the SSC loss energy rates.

Another important implication of this analysis is that the spectral
breaks associated with incomplete particle cooling can be
significantly different from the generally assumed $\Delta\alpha =
0.5$.  Even for synchrotron-dominated cooling, the SSC spectral break
is $\Delta\alpha = 1/4$.  If SSC cooling dominates, then a range of
spectral breaks can occur.  Their values will be related to the
particle injection indices, and consequently, to the spectral indices
of the lower energy emission.  For synchrotron emission, the spectral
break is $\Delta\alpha = (4-p)/2 = (3-2\alpha_l)/2$ where $\alpha_l$
is the index of the low energy (uncooled) part of the spectrum, while
for SSC emission, $\Delta\alpha = (4-p)/4 = (3-2\alpha_l)/4$.
However, because of the short energy loss time scales associated with
SSC cooling for typical blazar parameters and light travel time
effects through the plasma blob, detection of these spectral breaks
may be difficult (Chiaberge \& Ghisellini 1999; see also \S~4).  We
note that the spectral breaks we discuss here are distinct from those
which arise from energy cut-offs in the electron distribution
function.  Dermer et al.\ (1997) found that such breaks will be seen
if they are due to a low energy cut-off in the electron distribution
or if the observation times are sufficiently short for high energy
cut-offs.  For example, the lowest energy breaks in both the
synchrortron and SSC components shown in Fig.~\ref{example_seds}
reflect our choice of $\gamma_1 = 400$.

\section{Application to Blazar Observations}
As we have shown, even if the injected electron distribution varies by
a significant amount, dominant SSC cooling will produce synchrotron
emission with spectral index values close to $\alpha = 3/2$.  This is
precisely what has been observed for the $\gamma$-ray blazar 3C~279.
Hartman et al.\ (2001) have used the model of B\"ottcher et al.\
(1997) and B\"ottcher \& Bloom (2000) to fit the infrared through
$\gamma$-ray spectral energy distribution (SED) of 3C~279 for several
epochs during which simultaneous multiwavelength data were taken.  The
B\"ottcher et al.\ model calculates, in detail, the spectra and energy
losses due to synchrotron, SSC, and inverse Compton (IC) scattering of
accretion disk radiation both directly from the disk and rescattered
in the putative broad line region (BLR).  The SED fits of 3C~279,
which describe the optical/UV spectra fairly well, yield spectral
indices for the synchrotron emission in these wavebands of $\alpha
\simeq 1.4$--1.7 for injection indices of $p = 1.9$--3.1.  If
synchrotron cooling were the most important process, then we should
find a different range of indices, $\alpha = p/2 = 0.95$--1.55.

Although the optical/UV spectral shape for 3C~279 indicates that SSC
emission is the principal cooling mechanism, this does not require
that the observed emission in the gamma-ray band will be dominated by
SSC flux.  During the period in which 3C~279 was seen to have its
largest observed $\gamma$-ray flare (1996 Jan 30--Feb 6, epoch P5b of
Hartman et al.\ 2001), the observed $\gamma$-ray flux was found, using
the B\"ottcher et al.\ model, to be consistent with being mostly
composed of disk and BLR IC emission, not SSC emission.  However, the
greater relative strength of the disk and BLR IC components is due to
the fact that the assumed observer angle $\theta_{\rm obs} =2^\circ$
lies well within the Doppler beaming angle of $\theta_D \simeq
1/\Gamma = 1/13 = 4.4^\circ$ for their calculations.  As Dermer (1995)
pointed out, external Compton scattered emission has a substantially
narrower beaming pattern than SSC emission, and therefore it will have
a relative magnitude greater by at least a Doppler factor (${\cal D}
\sim 20$ in this case) as compared to SSC emission for the same total
luminosity.  For the calculations performed for epoch P5b, the SSC
component dominates the disk and BLR IC components for observing
angles $\theta_{\rm obs} \ga \theta_D$.  Consistent with this, the
energy loss rates calculated by the B\"ottcher et al.\ code do show
that SSC emission is the principal cooling mechanism.  Application of
this model to multiwavelength observations of the $\gamma$-ray blazar
PKS~0528+134 had much less success in reproducing the optical spectral
shape (Mukherjee et al.\ 1999).  The observed optical emission was
significantly harder than that of 3C~279 while the model parameters
used to fit the PKS~0528+134 SEDs were consistent with cooling
predominantly by SSC emission.  This suggests that some other cooling
mechanism controls the particle evolution, or as was proposed by
Mukherjee et al., that some additional processes, such as particle
re-acceleration downstream in the bulk outflow, are occurring.

BL Lac objects may provide a better ``laboratory'' in which to test
the importance of SSC cooling than the flat spectrum
\parbox{3.5in}{\epsfxsize=3.5in\epsfbox{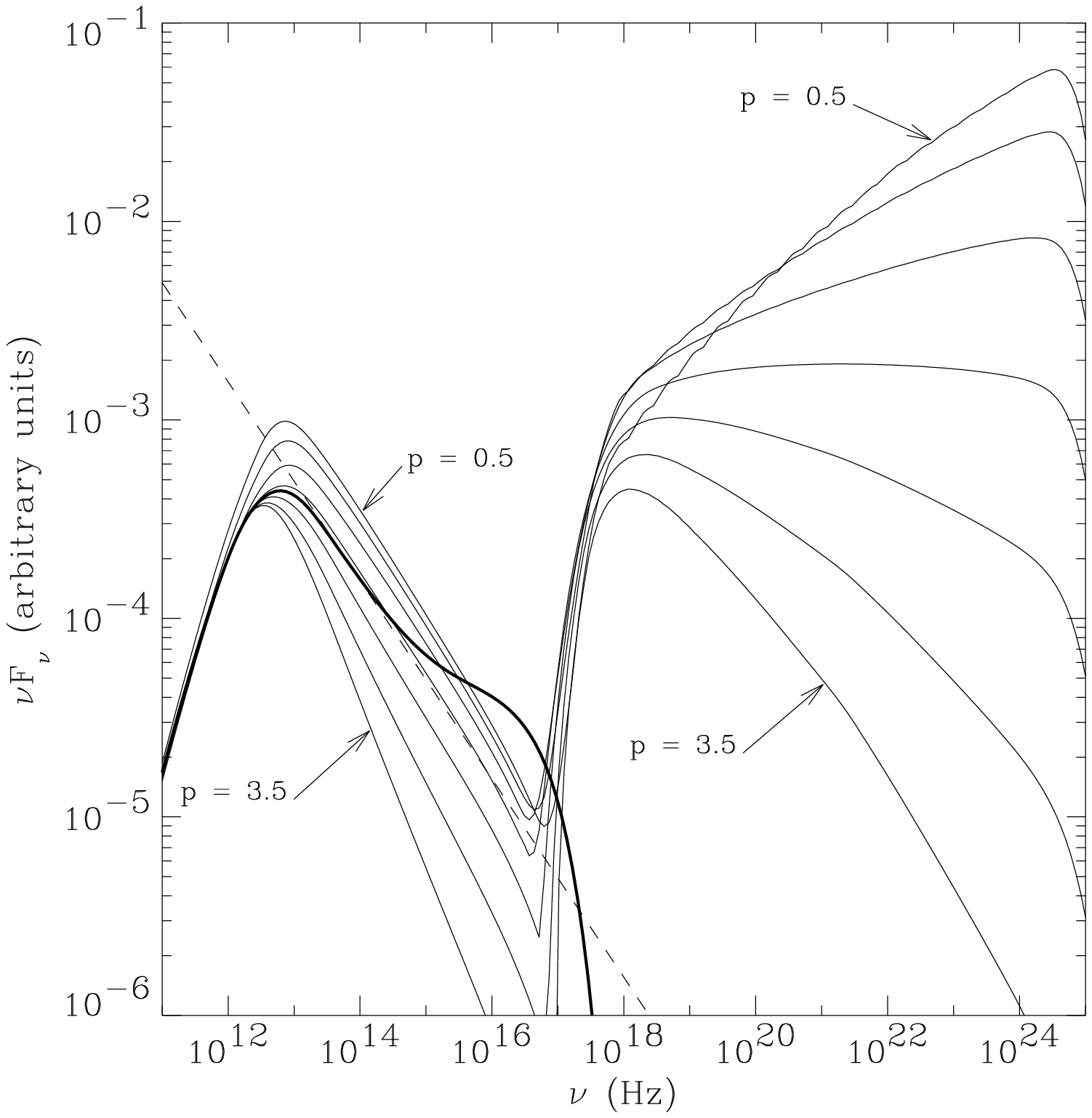}}
\figcaption{Time-averaged synchrotron and SSC spectra for values of
the particle injection index $p = 0.5$, 1, 1.5, 2, 2.5, 3., 3.5 (thin
curves).  These spectra have been computed for parameters similar to
those used for modeling the multiwaveband data of 3C~279 for epoch P5b
(Hartman et al.\ 2001): $\gamma_1 = 400$, $\gamma_2 = 10^5$, $B =
1.5$\,G, $R_b = 6 \times 10^{16}$\,cm, $n_e = 100$\,cm$^{-3}$.  The
thick curve is the time-averaged synchrotron spectrum for $p=2$, but
using the full Klein-Nishina cross-section to compute the SSC energy
loss rates (B\"ottcher et al.\ 1997).  The dashed line shows a $F_\nu
\propto \nu^{-3/2}$ spectrum.
\vspace{0.1in}
\label{example_seds}}
\noindent
radio quasars (FSRQs) as they generally do not show accretion disk
components or strong broad emission lines, and the high energy
emission can often be accounted for solely by SSC emission.  Since the
synchrotron and SSC components have the same Doppler beaming patterns,
comparing the observed luminosities of these two components will
indicate whether synchrotron or SSC cooling is more important.
Ironically, BL Lacertae itself does occasionally exhibit weak broad UV
emission lines, and Madejski et al.\ (1999) and B\"ottcher \& Bloom
(2000) have shown that a BLR IC component seems to be required to
explain the $\gamma$-ray emission.  As with PKS~0528+134, its optical
spectrum is not well modeled by synchrotron emission from an
SSC-cooled particle distribution.

This spectral diagnostic for SSC cooling need not be restricted to the
optical/UV bands.  For high frequency BL Lac objects (HBLs), the
synchrotron component peaks in the UV rather than in the infrared as
it does for low frequency BL Lac objects (LBLs) such as BL Lacertae
(Ghisellini et al.\ 1998).  In HBLs, the $\alpha \simeq 3/2$ signature
of SSC cooling will appear in the X-ray range rather than in the
optical/UV.  During the 1994 May X-ray/TeV flare of the HBL Mrk 421,
the X-ray energy spectral index was seen to evolve from $\alpha = 1.3$
during the rise of the flare to a maximum value of $\alpha = 1.5$ as
the flux diminished (Takahashi et al.\ 1996).  This spectral evolution
was attributed to synchrotron cooling, but the extremal value of
$\alpha = 1.5$ suggests that SSC cooling may be the primary mechanism.
Nearly simultaneous $\gamma$-ray fluxes measured at 100\,MeV through
TeV energies seem to indicate that the SSC luminosity was indeed
greater than the synchrotron luminosity during this flare episode
(Macomb et al.\ 1995).  By contrast, the HBL Mrk 501 does not seem to
be predominantly cooled by SSC emission.  The SEDs shown in Petry et
al.\ (2000) and Katarzy\'nski, Sol, \& Kus (2001) have a substantially
harder synchrotron spectral index than $\alpha = 3/2$, and the
synchrotron luminosity consistently exceeds the SSC luminosity.  The
HBL PKS~2155$-$304 exhibited soft X-ray spectral indices of $\alpha =
1.3$--1.7 during multiwavelength observations in 1994 May (Urry et
al.\ 1997). This suggests that SSC cooling was dominant during these
epochs.  Unfortunately, simultaneous $\gamma$-ray coverage was not
available, so direct comparisons of the SSC and synchrotron
luminosities cannot be made.

\section{Discussion}

If SSC emission is the primary energy loss mechanism, then in the
co-moving frame of the plasma blob, the synchrotron photon energy
density must exceed the energy densities associated with the other
relevant loss mechanisms, synchrotron emission and inverse Compton
scattering of external photons.  For SSC cooling to exceed synchrotron
cooling, this requires
\begin{equation}
\sigma_T R_b n_{\rm tot} \gamma_2^{3-p}\gamma_1^{p-1}\pfrac{p-1}{3-p} > 1,
\end{equation}
which follows from Eq.~\ref{usyn}. This relation can yield additional
constraints on the size scale and particle content of the emitting
plasma by combining it with upper limits on the blob radius $R_b$
given by variability time scales, inferred values of the particle
energy lower cut-off $\gamma_1$ from X-ray spectral shapes, and values
of the non-thermal particle density $n_{\rm tot}$ and magnetic field
$B$ from fits to the observed optical and X-ray spectra.

Comparing the synchrotron energy density to the energy densities of
external radiation is problematic because of uncertainties regarding
the amount of reprocessing of the central disk radiation the ambient
matter in the quasar environment can provide.  If the synchrotron
spectrum does show the signature of SSC cooling, then constraints on
the properties of this ambient matter follow.  Assuming it can be
modeled by a spherical distribution with characteristic radius
$\rblr$, and if the emitting plasma lies within this radius, then the
energy density of reprocessed disk radiation in the blob frame is
approximately\footnote{The location of the emitting plasma does not
appear in Eq.~\ref{ublr} since the energy density of reprocessed
radiation inside of a spherically symmetric shell of material which
reprocesses the flux from a central point source and re-emits it
isotropically is approximately uniform throughout the interior of the
shell.  This follows from a similar line of reasoning that the net
gravitational force inside a uniform shell of matter is zero
everywhere within the shell.}
\begin{eqnarray}
u_{\rm BLR} & \simeq & \Gamma^2 \tau_{\rm BLR} \frac{L_{\rm disk}}
                       {4\pi c r_{\rm BLR}^2}.\
\label{ublr}
\end{eqnarray}
Here $L_{\rm disk}$ is the disk luminosity, and $\tau_{\rm BLR}$ is
the characteristic optical depth for reprocessing the disk radiation.
Contributions to this optical depth may include Thomson scattering,
resonant line scattering, scattering by dust, and pure absorption
(with subsequent re-emission).  Combining this with the expression for
$\usyn$ (Eq.~\ref{usyn}) and reasonable estimates of $L_{\rm disk}$
and $\Gamma$, one can infer limits for the quantity $\tau_{\rm
BLR}/r_{\rm BLR}^2$.  For epoch P5b for 3C~279, Hartman et al.\ (2001)
used $\tau_{\rm BLR} = 0.003$, $r_{\rm BLR} \simeq 0.2$\,pc, $L_{\rm
disk} = 10^{46}$\,erg~s$^{-1}$, and parameters similar to those used
to produce Fig.~\ref{example_seds}.  These parameters yield
$\usyn/u_{\rm BLR} \sim 40$, confirming the dominance of SSC cooling.
A similar calculation can be applied to the direct disk radiation
(DS93), and for the 3C~279 epoch P5b model we find $\usyn/u_{\rm disk}
\sim 10$.

%In order for the direct disk radiation to be important, the height of
%the plasma blob above the disk must be comparable to the
%characteristic size of the disk emitting region, otherwise Doppler and
%Klein-Nishina effects will suppress the scattering of these photons
%(DS93).  At these heights, disk photons are incident upon the plasma
%at a wide range of angles, $\sim \pi$ to $\pi/2$, but the range of
%Doppler factors for the photons is relatively narrow, ${\cal D} \sim
%1/2\Gamma$ to $1/\Gamma$.  Since the transformation of the energy
%density from the stationary frame to the blob frame consists of a
%factor ${\cal D}^2$, we take $u_{\rm disk} \sim L_{\rm disk}/\Gamma^2
%4\pi c z_i^2$ as a crude estimate (see also DS93).  For the modeling
%of the epoch P5b data, Hartman et al.\ (2001) used $z_i = 0.025$\,pc,
%and we find $\usyn/u_{\rm disk} \sim 50$.  The detailed calculations
%using the B\"ottcher et al.\ code give $\usyn/u_{\rm disk} \sim 10$.

Although we have assumed that the blazar emission is produced in a
single blob of plasma, this analysis also describes the time steady
emission of a continuous jet of material in which the non-thermal
electrons are injected at the base of the jet. As noted by Hartman et
al.\ (2001), this is equivalent to a stream of discrete blobs.  Even
for a single blob, deviations from the time-averaged spectrum in the
optical and UV wavebands may only be detectable on fairly short time
scales.  For the 3C~279 epoch P5b parameters, the maximum energy of
the instantaneous synchrotron spectrum has swept through the optical
band after a time from the initial injection of $t_{\rm obs} =
(1+z)t_{\rm max}/{\cal D} \sim 5 \times 10^3$\,s as seen by a distant
observer (see Eq.~\ref{tmax_ssc}; here $z$ is the source redshift).
In order to characterize variations in the injection of non-thermal
particles or even to measure the spectral breaks associated with
partially cooled particle distributions, optical and UV spectra must
be taken on time scales shorter than $t_{\rm obs}$.  However, for Mrk
421, X-ray observations during the 1994 May flare were able to detect
the associated spectral evolution (Takahashi et al.\ 1996).  This
implies that the size scale and/or particle density of the emitting
plasma for this object were sufficiently small to yield relatively
long cooling times (see Eq.~\ref{tmax_ssc}).

Finally, we note that this analysis underscores the role of Doppler
beaming in assessing the contribution of various emission processes to
the overall energy balance of the emitting plasma.  Overly general
assumptions about the apparent luminosity of the various inferred
$\gamma$-ray components may lead one to conclude that either external
IC or SSC emission is the most important process for controlling the
evolution of the relativistic particles in the jet outflow, when in
fact the relative strength of these components is highly dependent on
observer orientation (Dermer 1995).  We have shown that the presence
(or absence) of a $F_\nu \sim \nu^{-3/2}$ synchrotron spectral shape
provides a much more reliable measure of these processes.

\acknowledgments

We thank Bob Hartman and the anonymous referee for helpful comments
that have improved this paper.  MB is supported by NASA through
Chandra Postdoctoral Fellowship Award 9-10007, issued by the Chandra
X-Ray Center, which is operated by the Smithsonian Astrophysical
Observatory for and on behalf of NASA under contract NAS 8-39073.

%\begin{figure}
%\centerline{\epsfxsize=5in\epsfbox{example_seds.eps}}
%\caption{Time-averaged synchrotron and SSC spectra for values of the
%power-law particle injection index $p = 0.5$, 1, 1.5, 2, 2.5, 3., 3.5
%(thin curves).  These spectra have been computed for parameters
%similar to those used for modeling the multiwaveband data of 3C~279
%for epoch P5b (Hartman et al.\ 2001): $\gamma_1 = 400$, $\gamma_2 =
%10^5$, $B = 1.5$\,G, $R_b = 6 \times 10^{16}$\,cm, $n_e =
%100$\,cm$^{-3}$.  The thick curve is the time-averaged synchrotron
%spectrum for $p=2$, but using the full Klein-Nishina cross-section to
%compute the SSC energy loss rates (B\"ottcher et al.\ 1997).  The
%dashed line shows a $F_\nu \propto \nu^{-3/2}$ spectrum.
%\label{example_seds}}
%\end{figure}

\end{document}